\documentclass[a4paper,twocolumn,11pt,
unpublished
]{quantumarticle}
\pdfoutput=1
\usepackage[utf8]{inputenc}
\usepackage[english]{babel}
\usepackage[T1]{fontenc}
\usepackage{amsmath,amssymb,amsthm}
\usepackage{hyperref}

\usepackage[numbers,sort&compress]{natbib}

\usepackage{tikz}
\usepackage{tikz,pgfplots,pgf}
\usepackage{circuitikz}
\usetikzlibrary{arrows,decorations.pathreplacing}

\definecolor{darkgray176}{RGB}{176,176,176}
\definecolor{lightgray204}{RGB}{204,204,204}
\definecolor{color1}{rgb}{0.0000,0.4470,0.7410}
\definecolor{color2}{rgb}{0.8500,0.3250,0.0980}
\definecolor{color3}{rgb}{0.9290,0.6940,0.1250}
\definecolor{color4}{rgb}{0.4940,0.1840,0.5560}
\definecolor{color5}{rgb}{0.4660,0.6740,0.1880}
\definecolor{color6}{rgb}{0.3010,0.7450,0.9330}
\definecolor{color7}{rgb}{0.6350,0.0780,0.1840}

\usepackage{lipsum}
\usepackage{enumerate}
\usepackage[ruled,vlined,linesnumbered]{algorithm2e}

\newtheorem{theorem}{Theorem}

\newcommand{\px}{\sigma_x}
\newcommand{\pz}{\sigma_z}
\newcommand{\py}{\sigma_y}

\renewcommand{\le}{\leqslant}
\renewcommand{\leq}{\leqslant}
\renewcommand{\ge}{\geqslant}
\renewcommand{\geq}{\geqslant}

\begin{document}

\title{Belief Propagation Decoding of 
Quantum LDPC Codes with Guided Decimation}

\author{Hanwen Yao}
\affiliation{Duke Quantum Center, Duke University, Durham, NC 27701, USA}
\affiliation{Department of Electrical and Computer Engineering, Duke University, Durham, NC 27708, USA}
\email{hanwen.yao@duke.edu}
\author{Waleed Abu Laban}
\affiliation{Department of Electrical Engineering, Chalmers University of Technology, Gothenburg, Sweden}
\email{laban@student.chalmers.se}
\author{Christian H\"ager}
\affiliation{Department of Electrical Engineering, Chalmers University of Technology, Gothenburg, Sweden}
\email{christian.haeger@chalmers.se}
\author{Alexandre Graell i Amat}
\affiliation{Department of Electrical Engineering, Chalmers University of Technology, Gothenburg, Sweden}
\email{alexandre.graell@chalmers.se}
\author{Henry D. Pfister}
\affiliation{Duke Quantum Center, Duke University, Durham, NC 27701, USA}
\affiliation{Department of Electrical and Computer Engineering, Duke University, Durham, NC 27708, USA}
\affiliation{Department of Mathematics, Duke University, Durham, NC, USA}
\email{henry.pfister@duke.edu}

\maketitle

\begin{abstract}
Quantum low-density parity-check (QLDPC) codes 
have emerged as a promising technique 
for quantum error correction.
A variety of decoders have been proposed 
for QLDPC codes and many of them utilize 
belief propagation (BP) decoding in some fashion. 
However, the use of BP decoding 
for degenerate QLDPC codes 
is known to have issues with convergence. 
These issues are typically attributed to 
short cycles in the Tanner graph and code degeneracy 
(i.e. multiple error patterns with the same syndrome). 
Although various methods have been proposed 
to mitigate the non-convergence issue, 
such as BP with ordered statistics decoding (BP-OSD) 
and BP with stabilizer inactivation (BP-SI), 
achieving better performance with lower complexity 
remains an active area of research.

In this work, we propose a decoder for QLDPC codes 
based on BP guided decimation (BPGD), 
which has been previously studied 
for constraint satisfaction and lossy compression problems. 
The decimation process is applicable to both 
binary and quaternary BP and it involves
sequentially fixing the value of the most reliable qubits to encourage BP convergence.
Despite its simplicity, We find that BPGD significantly reduces 
the BP failure rate due to non-convergence,
achieving performance on par with BP with ordered statistics decoding and BP with stabilizer inactivation, without the need to solve systems of linear equations. 
\end{abstract}

\section{Introduction}
In a quantum computing system, 
error correction is an essential building block 
to protect fragile quantum information against 
decoherence and other noise sources. 
The general framework of quantum stabilizer codes 
has been studied extensively \cite{calderbank1997quantum, calderbank1998quantum, gottesman1997stabilizer}. 
Using this framework, 
various quantum error-correcting codes 
have been constructed over the past two decades 
including toric codes \cite{kitaev1997quantum,kitaev2003fault}, 
surface codes \cite{dennis2002topological, bravyi2010tradeoffs, fowler2012surface, horsman2012surface}, 
and various quantum low-density parity-check (QLDPC) codes \cite{mackay2004sparse, tillich2013quantum, kovalev2012improved, kovalev2013quantum, haah2011local, Panteleev2021degeneratequantum, yang2023quantum,breuckmann2021quantum}. 
Among them, QLDPC codes provide a promising direction 
because they support multiple logical qubits 
and their low-weight stabilizers allow reliable syndrome measurement in practice. 
Until recently, researchers did not know how to construct QLDPC codes with constant rate and linear distance.
But, in a series of recent advances 
\cite{hastings2021fiber, breuckmann2021balanced, panteleev2021quantum, panteleev2022asymptotically},
researchers have overcome this obstacle and there are now constructions of asymptotically \emph{good} quantum LDPC codes 
with constant rate and linear minimum distance.

For classical communication, 
low-density parity-check (LDPC) codes 
were first proposed by Gallager \cite{gallager1962low} in 1962 and are now widely applied in practical communication systems \cite{DVB-S2,richardson2018design}.
LDPC codes are typically 
decoded with the belief propagation (BP) algorithm \cite{kschischang2001factor}, 
which has low complexity and can provide good performance for code rates close to the channel 
capacity~\cite{luby2001efficient,luby2001improved,richardson2001capacity,richardson2001design}. 
However, in the quantum scenario, the performance of BP decoding based on syndrome measurement is hindered by cycles in the Tanner graph and code degeneracy \cite{poulin2008iterative,babar2015fifteen,raveendran2021trapping}.
The first problem follows from the commutativity constraint imposed on stabilizer codes which produces
unavoidable cycles in the Tanner graph that degrade the BP decoding performance \cite{poulin2008iterative,babar2015fifteen}. 
The second is because good QLDPC codes are recognized to be highly \emph{degenerate}; 
this means that their minimum distance is much larger than the minimum weight of their stabilizers.
Due to degeneracy, the syndrome (i.e., stabilizer measurements) can be used to identify a correction procedure that works with high probability even when the actual error is not uniquely identified.
This hinders convergence of the BP decoding process \cite{poulin2008iterative,raveendran2021trapping}.
Since the initial application of BP to decode QLDPC codes by Poulin and Chung \cite{poulin2008iterative}, significant efforts have been made to enhance its performance. 
Various methods have been proposed to modify the BP decoding process itself, including random perturbation \cite{poulin2008iterative}, enhanced feedback \cite{wang2012enhanced}, grouping check nodes as super nodes \cite{babar2015fifteen}, parity-check matrix augmentation \cite{rigby2019modified}, neural BP \cite{liu2019neural,xiao2019neural,miao2023neural}, generalized BP \cite{old2023generalized}, adaptive BP with memory \cite{kuo2022exploiting}, 
and BP with trapping set dynamics \cite{chytas2024enhanced}. 
Alternatively, post-processing methods have also been explored to improve performance.
In \cite{Panteleev2021degeneratequantum}, when BP fails to converge, 
it was proposed to 
use ordered statistics decoding (OSD) to construct a syndrome-matching error pattern based on the soft information provided by BP. This is called the BP-OSD algorithm.
Another post-processing approach introduced in \cite{du2022stabilizer} involves iteratively running BP with stabilizer inactivation (BP-SI). 

In this work, we improve the BP decoding performance for QLDPC codes by combining it with guided decimation. 
The term ``decimation'' refers to the process of sequentially fixing variables to hard decisions during iterative decoding~\cite{mezard2002analytic,montanari2007solving}. 
To motivate BP guided decimation (BPGD), 
we propose a sampling decoding approach for QLDPC codes which 
samples an error pattern based on their posterior probabilities.
Then we explain how BPGD can be used to approximate the sampling decoder.
In the proposed BPGD algorithm, we incorporate the decimation method by iteratively
fixing the most reliable qubit based on its marginals estimated by BP. 
Despite its simplicity, we show that BPGD achieves performance 
on par with both order-0 BP-OSD and BP-SI. 
Notably, BPGD exhibits lower complexity than BP-OSD and comparable complexity to BP-SI, without the need to solve any systems of linear equations. 
Additionally, through a randomized version of BPGD, we shed light on how code degeneracy contributes to the non-convergence issue in syndrome BP decoding over QLDPC codes. 
Our experiment also shows that guided decimation improves BP convergence, and highlights how BPGD benefits from code degeneracy.
Furthermore, we extend BPGD from binary symbols to quaternary symbols, demonstrating competitive performance compared to BP-OSD in the high error rate regime over depolarizing noise.

\section{Preliminaries}
\label{sec:prelim}
\subsection{Binary Linear Codes}
Let $\mathbb{F}_2 = \{0,1\}$ be the binary Galois field 
defined by modulo-2 addition and multiplication.
A length-$n$ binary linear code $\mathcal{C}\subseteq\mathbb{F}_2^n$ 
is a subset of length-$n$ binary strings satisfying 
$w+x \in \mathcal{C}$ for all $w,x \in \mathcal{C}$.
Such a code forms a vector space over $\mathbb{F}_2$.
A generator matrix $G\in \mathbb{F}_2^{k \times n}$ for 
$\mathcal{C}$ is a $k \times n$ matrix whose rows span the code.
A parity-check matrix $H\in \mathbb{F}_2^{(n-k) \times n}$ 
for $\mathcal{C}$ is an $(n-k) \times n$ matrix 
whose rows are orthogonal to the code.

For classical communication where the codeword $x\in \mathcal{C}$ 
is corrupted by an additive error vector $z \in \mathbb{F}_2^n$, 
the received vector is given by $y=x+z$.
Then, the optimal recovery process for $x$ given $y$ is 
choosing a codeword $\widehat{x}$ that maximizes the conditional probability given $y$.
Since $Hx = 0$, it follows that the syndrome vector 
$s \triangleq Hy \in \mathbb{F}_2^{n-k}$ satisfies
\begin{equation}
s= H(x+z) = Hz.
\end{equation}
For the binary symmetric channel (BSC) where $z$ is an i.i.d.\ vector, the optimal recovery process can also be implemented by \emph{syndrome decoding} where the syndrome $s$ is mapped to the minimum-weight error vector in the coset $\{ u \in \mathbb{F}_2^{n} \, |\, Hu = s \}$ containing $z$.
For more details on this well-known classical setup, see~\cite{Blahut-2003}.

\subsection{Stabilizer Formalism}
An $[[n,k]]$ quantum stabilizer code is an error correction code designed to protect $k$ logical qubits with $n$ physical qubits against noise.
We first define the Pauli operators to establish the definition of quantum stabilizer codes.
For a single qubit, 
its pure quantum state is represented as a unit vector
in the two-dimensional Hilbert space $\mathbb{C}_2$.
The Pauli operators for a single qubit system are defined
as the $2\times 2$ complex Hermitian matrices
\begin{equation}
\begin{aligned}
    &I = \begin{bmatrix}
        1 & 0 \\ 0 & 1
    \end{bmatrix},\,
    \px = \begin{bmatrix}
        0 & 1 \\ 1 & 0
    \end{bmatrix},\\
    &\pz = \begin{bmatrix}
        1 & 0 \\ 0 & -1
    \end{bmatrix},\,
    \py = \imath\sigma_x \sigma_z = \begin{bmatrix}
        0 & -\imath\\ \imath & 0
    \end{bmatrix},
\end{aligned}
\end{equation}
where $\imath=\sqrt{-1}$,
and they form a basis for all $2\times 2$ complex matrices.
For an $n$-qubit system, we are working in the $n$-fold Kronecker product of the two-dimensional Hilbert space 
$\mathbb{C}_2^{\otimes n}$. 
Given two length-$n$ binary vectors $a=(a_1,a_2,\ldots,a_n)\in \mathbb{F}_2^n$, and $b=(b_1,b_2,\ldots,b_n)\in \mathbb{F}_2^n$, 
we define the $n$-fold Pauli operator $D(a,b)$ as
\begin{equation}
    D(a,b) = \px^{a_1}\pz^{b_1}\otimes \px^{a_2}\pz^{b_2}\otimes \ldots \otimes \px^{a_n}\pz^{b_n}.
\end{equation}
It follows that the Pauli operators $\imath^k D(a,b)$ with $a,b\in\mathbb{F}_2^n$ and an overall phase $\imath^k$ with $k\in\{0,1,2,3\}$ form the $n$-qubit Pauli group, denoted by $\mathcal{P}_n$, with the multiplication rule
\begin{equation}
\begin{aligned}
    D(a,b)D(a',b') 
    &= (-1)^{a'b^T}D(a+a',b+b') \\
    &= (-1)^{a'b^T + b'a^T}D(a',b')D(a,b).
\end{aligned}
\end{equation}
The \emph{symplectic inner product} between length-$2n$ binary vectors $(a,b)$ and $(a',b')$ is defined by
\begin{equation}
\begin{aligned}
    \langle (a,b),(a',b') \rangle_s 
    &\triangleq (a',b') \Lambda (a,b)^T \bmod 2
    \\ &= b'a^T + a'b^T \bmod 2,
\end{aligned}
\end{equation}
where
\begin{equation}
    \Lambda = \begin{bmatrix} 0 & I_n \\ I_n & 0 \end{bmatrix}.
\end{equation}
This equals 0 if the two Pauli operators $D(a,b)$ and $D(a',b')$ commute and 1 if they anti-commute.

A \emph{quantum stabilizer code} $\mathcal{C}$ with $n$ physical qubits 
is defined by a commutative subgroup $\mathcal{S}\subseteq \mathcal{P}_n$ with $-I_2^{\otimes n}\notin \mathcal{S}$. 
The subgroup $\mathcal{S}$ is referred to as the \emph{stabilizer group}, and the Pauli operators in $\mathcal{S}$ are called the \emph{stabilizers}.
The code space consists of all states in
$\mathbb{C}_2^{\otimes n}$ stabilized by $\mathcal{S}$:
\begin{equation}
    \mathcal{C} = \{|\psi\rangle\in\mathcal{C}_2^{\otimes n}:
    M|\psi\rangle = |\psi\rangle,\, \forall M\in\mathcal{S}\}.
\end{equation}
In other words, $\mathcal{C}$ consists of states that are +1 eigenstates of all the stabilizers in $\mathcal{S}$.
If $\mathcal{S}$ has $n-k$ independent generators, 
the code space has dimension $k$.

The \emph{weight} of a Pauli operator in $\mathcal{P}_n$
is defined to be the number of elements in its $n$-fold Kronecker product that are not equal to $I$.
The \emph{distance} of a stabilizer code $\mathcal{C}$ is defined as the minimum weight of all Pauli operators in $N(\mathcal{S})\backslash \mathcal{S}$, 
where $N(\mathcal{S})$ denotes the normalizer group of $\mathcal{S}$ in $\mathcal{P}$.
If code $\mathcal{C}$ has distance $d$, we call $\mathcal{C}$ an $[[n,k,d]]$ quantum stabilizer code.
In particular, $\mathcal{C}$ is called \emph{degenerate} if its distance $d$ is larger than the minimum weight of its stabilizers.

In the symplectic representation, the stabilizer group $\mathcal{S}$ is constructed from the rows of the stabilizer matrix 
\begin{equation}
H = [H_x,H_z],
\end{equation}
where $H_x,H_z\in\mathbb{F}_2^{m\times n}$ are binary matrices with $m$ rows and $n$ columns.
In particular, each row $(h_x,h_z)$ of $H$ defines the stabilizer $D(h_x,h_z)$ and the set of stabilizers defined by all rows generates the stabilizer group $\mathcal{S}$.
In this way, the constraint requiring all the stabilizers in $H$  to commute with each other can be expressed as 
\begin{equation}
H \Lambda H^T = H_x H_z^T + H_z H_x^T = 0.
\end{equation}
There is an important class of stabilizer codes, known as 
Calderbank–Shor–Steane (CSS) codes \cite{calderbank1996good,steane1996multiple}, where each stabilizer has the form $D(a,0)$ (i.e., only $\px$ operators) or $D(0,b)$ (i.e., only $\pz$ operators).
In this case, following the convention in \cite{calderbank1996good}, we have
\begin{equation}
H_x = \begin{bmatrix} 0 \\ G_2 \end{bmatrix}, \quad H_z = \begin{bmatrix} H_1 \\ 0 \end{bmatrix},
\label{eq:H1_G2}
\end{equation}
where $H_1 \in \mathbb{F}_2^{(n-k_1)\times n}$ is the parity-check matrix of a classical $[n,k_1]$ code $\mathcal{C}_1$ and $G_2 \in \mathbb{F}_2^{k_2\times n}$ is the generator matrix of a classical $[n,k_2]$ code $\mathcal{C}_2$.
Thus, the stabilizer matrix has the form
\begin{equation}
    H = \begin{bmatrix}
        0 & H_1 \\
        G_2 & 0
    \end{bmatrix}.
\end{equation}
For CSS codes, the commutativity constraint for the stabilizers given by $H \Lambda H^T = 0$ reduces to $G_2 H_1^T = 0$, which is equivalent to $\mathcal{C}_2 \subseteq \mathcal{C}_1$. 
Throughout this work, we will focus our discussion on the CSS codes.

\subsection{Syndrome Decoding of Stabilizer Codes}
\label{subsec:syndrome_decoding}
In this work, we consider two types of code capacity noise models: 
the bit-flip noise model and the depolarizing noise model.
In both of these models, the encoded state $|\psi\rangle$ 
of an $[[n,k]]$ stabilizer code $\mathcal{C}$ is corrupted by 
an $n$-qubit Pauli error $E=D(x,z)\in\mathcal{P}_n$ as $|\psi\rangle\rightarrow E|\psi\rangle$. 
The goal of the decoder is to detect and correct this error by conducting measurements on all the stabilizers in the parity-check matrix $H$.

Measuring the stabilizer $M = D(a,b) \in \mathcal{S}$ reveals the symplectic inner product $\langle (a,b),(x,z)\rangle_s$, indicating whether $E$ commutes or anti-commutes with $M$.
When measuring all the stabilizers in $H$, the result can be expressed as a length-$m$ binary syndrome vector 
\begin{equation}
s = (x,z) \Lambda H^T.
\end{equation}
In this work, we assume that all 
syndrome measurements are uncorrupted (i.e. do not contain errors).

For CSS codes, $H_1$ in $H_x$ defines $\pz$-stabilizers that interact with Pauli-$\px$ errors and $G_2$ in $H_z$ defines $\px$-stabilizers that interact with Pauli-$\pz$ errors.
Thus, we can separate the length $m=n-k_1+k_2$ syndrome vector into two parts $s = (s_x,s_z)$ where $s_x = x H_1^T$ has length $n-k_1$ and $s_z = z G_2^T$ has length $k_2$.

Based on its syndrome, the Pauli error $E$ that affects 
the qubits can be categorized as follows:
\begin{enumerate}
    \item {\bf Detectable Error}:
    $E$ is called \emph{detectable} if its 
    syndrome $s$ is not the all-zero vector. 
    In other words, $E$ is detectable if it anticommutes 
    with at least one of the stabilizers in $\mathcal{S}$.
    Otherwise, $E$ is called \emph{undetectable}.
    \item {\bf Degenerate Error}:
    If $E$ is undetectable, it is called \emph{degenerate} 
    if it belongs to the stabilizer group, 
    i.e., $E\in\mathcal{S}$.
    In this case, $E$ preserves the encoded state and needs no correction.
    \item {\bf Logical Error}:
    If $E$ is undetectable and it does not belong
    to the stabilizer group, i.e., 
    $E\in N(\mathcal{S})\backslash\mathcal{S}$,
    it is called a \emph{logical} error, 
    and it alters the logical state of the encoded qubits.
\end{enumerate}

After obtaining the syndrome $s$ through measurement, 
the decoder aims to find an estimated $\widehat{E}$ 
with high posterior probability yielding this syndrome.
Then, a reverse operator $\widehat{E}^{\dag}$, which in our case is equal to $\widehat{E}$ since $\widehat{E}$ is Pauli, 
can be applied to the affected qubits as 
$E|\psi\rangle\rightarrow \widehat{E}E|\psi\rangle$ 
for error correction.
This decoding process has the following four possible outcomes:
\begin{enumerate}
    \item {\bf Failure}: 
    The decoder fails to provide an estimated 
    $\widehat{E}$ that yields the syndrome $s$.
    \item {\bf Successful (Exact Match)}:
    The estimated error is equal to the channel error, 
    i.e., $\widehat{E} = E$.
    \item {\bf Successful (Degenerate Error)}:
    The difference between the estimated error and the channel error 
    is a degenerate error, i.e., $\widehat{E}E\in\mathcal{S}$.
    \item {\bf Failure (Logical Error)}: 
    The difference between the estimated error and the channel error
    is a logical error, 
    i.e., $\widehat{E}E\in N(\mathcal{S})\backslash\mathcal{S}$.
\end{enumerate}

Since there are $2^m$ possible syndromes, using a look-up table for error estimation quickly becomes infeasible as the number of qubits increases because the number of stabilizers $m$ in $H$ typically scales linearly with $n$.
Thus, a computationally efficient decoder is required for long stabilizer codes.

Using the terminology 
from~\cite{iyer2015hardness}, 
here we describe two decoding strategies for stabilizer codes.
\paragraph{Quantum Maximum Likelihood (QML) Decoding} 
This decoding strategy aims at finding 
the most probable error $E = D(X,Z)$ given the syndrome, where $X,Z \in \mathbb{F}_2^n$ are random vectors drawn from some Pauli error distribution.
In this model, the syndrome is a random vector defined by $S=(X,Z) \Lambda H^T$. 
Given the observed syndrome event $S=s$, the decoder aims to find $\widehat{E}=D(\widehat{X},\widehat{Z})$ where 
\begin{multline}
(\widehat{X},\widehat{Z}) = \\
\underset{(x',z')\in \mathbb{F}_2^n \times \mathbb{F}_2^n}{\arg\max}\; \Pr \big((X,Z)=(x',z')\,|\,S=s \big).
\label{eq:qmld}
\end{multline}
For both the bit-flip noise model and
the depolarizing noise model, 
solving \eqref{eq:qmld} is at least as hard as the maximum-likelihood decoding problem in classical coding theory, which is well-known to be NP-complete \cite{berlekamp1978inherent}.

\paragraph{Degenerate Quantum Maximum Likelihood (DQML) 
Decoding} 
Denote the coset of the stabilizer group $\mathcal{S}$ shifted by a Pauli error $E\in\mathcal{P}_n$ as $E\mathcal{S}$, 
and denote the quotient group of all cosets of $\mathcal{S}$ in $\mathcal{P}_n$ as $\mathcal{P}_n/\mathcal{S}$.
In the context of a stabilizer code, where all Pauli errors in the same coset $E\mathcal{S}$ affect the logic state of encoded qubits 
equivalently, the optimal decoding strategy would be finding the most probable coset given the syndrome event $S = s$ as
\begin{multline}
\widehat{E\mathcal{S}} = \\
\underset{E\mathcal{S} \in \mathcal{P}_n/\mathcal{S}}{\arg\max}\; 
\hspace{-1mm}
\sum_{D(x',z')\in E\mathcal{S}}
\hspace{-1mm}
\Pr\big((X,Z)=(x',z')|S=s \big).
\label{eq:dqmld}
\end{multline}
After that, the decoder can choose one error pattern $\widehat{E}$ in coset $\widehat{E\mathcal{S}}$ as the error to be corrected, as the specific error selected does not affect the result due to degeneracy.
This decoding problem has been shown to be $\#$P-complete \cite{iyer2015hardness}, 
which is computationally much harder than the 
QML decoding problem.

\subsection{CSS Codes with Bit-Flip Errors}
\label{subsec:css_bitflip}
In Section~\ref{sec:BP}, \ref{sec:BPGD}, and \ref{sec:BPGD_rd} 
that follow, we will focus on the syndrome decoding problem of CSS codes 
with the bit-flip errors, where each encoded qubit is independently affected by a Pauli-$\px$ error with probability $p_x$. 
In this model, the Pauli error has the form $E=D(X,0)$, with $X\in\mathbb{F}_2^n$ being a random vector following the distribution:
\begin{equation} 
\label{eq:error_dist}
    \Pr \big(X=(x_1,x_2,\ldots,x_n)\big) = \prod_{i=1}^n p_x^{x_i}(1-p_x)^{1-x_i}
\end{equation}
In the following, we will use the random vector $X$ to denote the channel error for simplicity. 
For CSS codes in this case, the decoder only needs to consider part of the syndrome $S_x=XH_1^T$, where $H_1$ is the submatrix of $H_z$ as in \eqref{eq:H1_G2}.
The QML decoding problem then becomes finding $\widehat{X}$ where 
\begin{equation}
\begin{aligned}
\widehat{X} 
&= \underset{x\in\mathbb{F}_2^n}{\arg\max}\; \Pr(X=x\,|\,S_x=s_x)\\
&= \underset{x\in\mathbb{F}_2^n\,:\,xH_1^T = s_x}{\arg\min}\; \mathrm{wt}(x)
\end{aligned}
\end{equation}
This is equivalent to the maximum-likelihood decoding problem for the BSC in classical coding theory, 
which is known to be NP-complete~\cite{berlekamp1978inherent}.
Therefore, we turn to the low-complexity
belief propagation (BP) algorithm.

\section{Belief Propagation Decoding}
\label{sec:BP}
BP is a low-complexity algorithm that solves inference problems for graphical models.
Originated in the 1960s, Gallager's LDPC decoder \cite{gallager1962low} already contains the essence of BP, 
which was later formalized by Pearl in \cite{pearl1982reverend,pearl1988probabilistic}.
In 1993, the discovery of turbo codes \cite{berrou1993near} brought along the turbo decoding algorithm, 
which is now recognized as an instance of BP \cite{mceliece1998turbo}.
In the late 1990s, those ideas led to the formalization of factor graphs and the sum-product algorithm 
\cite{kschischang2001factor,aji2000generalized}.
Some strengths of BP decoding are that it is highly parallelizable and it can achieve near-optimal decoding of well-designed LDPC codes and turbo codes.
It also has wide applications in other problems on graphical models, such as the spin glass models \cite{mezard2001bethe},
lossy compression with the low-density generator matrix codes \cite{filler2007binary,aref2015approaching}, 
and constraint satisfaction problems such as $K$-SAT \cite{montanari2007solving,coja2011belief}.

BP was first introduced to the quantum decoding problem by Poulin and Chung in \cite{poulin2008iterative}.
However, the performance of BP for the quantum syndrome decoding problem for the QLDPC codes 
is degraded due to degeneracy \cite{poulin2008iterative,raveendran2021trapping}.
In this section, 
we first review the binary BP algorithm for syndrome decoding 
of stabilizer codes over bit-flip errors.
Then, we discuss prior improvements for BP that mitigate the non-convergence problem, 
including BP-OSD \cite{Panteleev2021degeneratequantum} 
and BP-SI \cite{du2022stabilizer}.

\subsection{Belief Propagation for Bit-Flip Errors}
\label{subsec:syndrome_bp}
Consider the setup described in Section~\ref{subsec:css_bitflip}
for the syndrome decoding problem of an 
$[[n,k]]$ CSS code over bit-flip noise. 
Denote the Pauli-$\px$ error as a random vector 
$X = (X_1,X_2,\ldots,X_n)\in\mathbb{F}_2^n$
following the distribution in \eqref{eq:error_dist}.
BP is an iterative message-passing algorithm 
runs on a Tanner graph representing $H_1$ 
that provides an estimate of the marginal probability 
\begin{equation}
\Pr(X_i = x_i\,|\,S_x = s_x)
\end{equation}
for all $i \in [n]$ and $x_i\in\{0,1\}$.

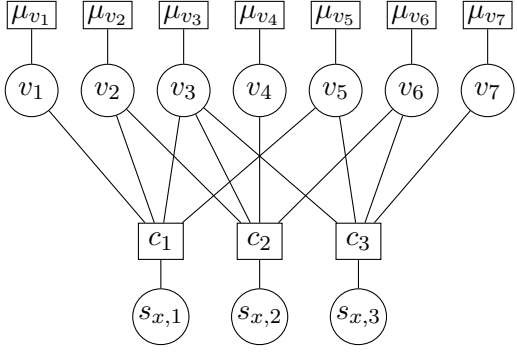
\begin{figure}[t]
    \centering
    \begin{tikzpicture}
    \foreach \x in {1,2,...,7}
    {
      \node[draw, circle, minimum size=0.3cm, inner sep=0.1cm] (q\x) at (\x, 0) {$v_\x$};
      \node[draw, inner sep=0.05cm] (cc\x) at (\x, 1) {$\mu_{v_{\x}}$};
      \draw (q\x) -- (cc\x);
    }
      
    \foreach \x/\y in {2.7/1,4/2,5.3/3}
    {
      \node[draw] (c\y) at (\x, -2) {$c_\y$};
      \node[draw, circle, minimum size=0.3cm, inner sep=0.03cm] (s\y) at (\x, -3) {$s_{x,\y}$};
      \draw (c\y) -- (s\y);
    }
    
    \foreach \x in {1,2,3,5}
      \draw (q\x) -- (c1);
    \foreach \x in {2,3,4,6}
      \draw (q\x) -- (c2);
    \foreach \x in {3,5,6,7}
      \draw (q\x) -- (c3);
    \end{tikzpicture}
    \caption{Tanner graph of $H_1$ for the $[[7,1,3]]$ Steane code 
    }
    \label{fig:Steane}
\end{figure}

Let $H_1$ be the $\pz$-stabilizer matrix in \eqref{eq:H1_G2}
with $m_1 = n-k_1$ rows and $n$ columns.
the Tanner graph $G = (V,C,E)$ is a bipartite graph with 
the variable nodes in $V=\{v_1,\ldots,v_n\}$ representing 
the elements of $X=(X_1,\ldots,X_n)$ and the check nodes in 
$C=\{c_1,\ldots,c_m\}$ representing the $\pz$-stabilizers in $H_1$.
A variable node $v_i$ is connected to a check node $c_j$ 
if $H_1 (i,j)=1$.
In addition, each variable node $v_i$ is connected to a degree 1 check node that represents the source of the bit-flip error, and each check node $c_j$ is connected to a degree 1 variable node that represents the source of a syndrome bit. 
An example Tanner graph representing $H_1$ for the $[[7,1,3]]$ Steane code \cite{steane1996multiple} with 
\begin{equation}
H_1 = \begin{bmatrix}
    1 & 1 & 1 & 0 & 1 & 0 & 0 \\
    0 & 1 & 1 & 1 & 0 & 1 & 0 \\
    0 & 0 & 1 & 0 & 1 & 1 & 1
\end{bmatrix}
\end{equation}
is shown in Fig.~\ref{fig:Steane}.

Now we describe the BP algorithm that runs on this Tanner graph.
Let $\mu_{v_i}$ denote the channel log-likelihood ratio (LLR) for $v_i$ as
\begin{equation}
\mu_{v_i} = \log\frac{\Pr(X_i=0)}{\Pr(X_i=1)} = \log\frac{1-p_x}{p_x}.
\label{eq:channel_message}
\end{equation}
This is the initial estimate of the overall LLR of $X_i$.
At iteration $t = 0$, 
set the message from each variable node $v_i$ to all its connected check node $c_j$ as
\begin{equation}
m^{(0)}_{v_i\rightarrow c_j} = \mu_{v_i}.
\end{equation}

Then, for iteration $t=0,1,\ldots,T$, given the syndrome $s_x = (s_{x,1},\ldots,s_{x,m})$, 
the messages between variable nodes and check nodes are updated as
\begin{equation}
m^{(t)}_{c_j\rightarrow v_i} = 
(-1)^{s_{x,j}} \tanh^{-1} 
\left(\tanh
\hspace{-2.5mm}
\prod_{v_k\in \partial c_j\backslash v_i}
\hspace{-2.5mm}
m^{(t)}_{v_k\rightarrow c_j}\right),
\label{eq:check_node_update}
\end{equation}
and 
\begin{equation}
m^{(t+1)}_{v_i\rightarrow c_j} = \mu_{v_i} + 
\sum_{c_k\in \partial v_i\backslash c_j}m^{(t)}_{c_k\rightarrow v_i},
\label{eq:bit_node_update}
\end{equation}
where $s_{x,j} \in\{0,1\}$ is the syndrome of the $\pz$-stabilizer corresponding to $c_j$ (i.e., defined by the $j$-th row of $H_1$),
$\partial c_j$ denotes the set of variable nodes in $V$ connected to $c_j$, and $\partial v_i$ denotes the set of check nodes in $C$ connected to $v_i$. 
This update rule is called the sum-product algorithm \cite{kschischang2001factor}.

In our implementation of BP, we also 
saturate the message to a domain $[-K,K]$ for numerical stability, 
where $K$ is a large fixed number.
Specifically, after every BP iteration, we clip the message 
$m^{(t+1)}_{v_i\rightarrow c_j}$ as
\begin{equation}
m^{(t+1)}_{v_i\rightarrow c_j} = 
\begin{cases}
K & 
\text{if } m^{(t+1)}_{v_i\rightarrow c_j} > K \\
m^{(t+1)}_{v_i\rightarrow c_j} & 
\text{if } m^{(t+1)}_{v_i\rightarrow c_j} \in [-K,K] \\
-K & 
\text{if } m^{(t+1)}_{v_i\rightarrow c_j} < -K
\end{cases}
\end{equation}
for all edges in the Tanner graph.
BP with saturation is commonly applied in practice to upper bound the soft information on the Tanner graph, and it plays a role in the error floor phenomenon and the stability of density evolution \cite{zhang2012will,schlegel2010dynamics,butler2014error,kudekar2014effect}.
In the software implementation for all our simulation results 
shown in this work, we set $K = 25$.
It is known that quantization and saturation for BP are complicated subjects, and we have yet to fully optimize for them.

The \emph{bias} for variable node $v_i$ after $t$ iterations is defined to be 
\begin{equation}
m^{(t)}_{v_i} = \mu_{v_i} + \sum_{c_k\in \partial v_i}m^{(t)}_{c_k\rightarrow v_i}.
\label{eq:bias}
\end{equation}

For sufficiently large $t$, one approximates the log-likelihood ratio of the marginal probabilities for $X_i$ with
\begin{equation}
m^{(t)}_{v_i} \approx \log\frac{\Pr(X_i=0\,|\,S_x = s_x)}{\Pr(X_i=1\,|\,S_x = s_x)}.
\label{eq:marginal}
\end{equation}
This approximation is exact if the Tanner graph $G$ is a tree \cite{kschischang2001factor}.
The sign of the bias $m^{(t)}_{v_i}$ represents the hard value toward which this variable node $v_i$ is biased, and the absolute value of the bias is denoted by
\begin{equation}
\gamma^{(t)}(v_i) = |m^{(t)}_{v_i}|,
\label{eq:gamma}
\end{equation}
or just $\gamma(v_i)$ if $t$ is clear from the context. This represents the \emph{reliability} of this variable node.
The larger the reliability, the more certain we are about its indicated hard value.

BP is commonly equipped with the following termination rule upon convergence \cite{poulin2008iterative}. 
After $t$ iterations, 
the estimated hard value for $X_i$ can be computed as
\begin{equation}
\widehat{x}^{(t)}_i = \begin{cases}
    0, & m^{(t)}_{v_i} > 0 \\
    1, & m^{(t)}_{v_i} \le 0
\end{cases}.
\end{equation}
The BP algorithm terminates if the computed hard values 
\begin{equation}
\widehat{x}^{(t)} = 
(\widehat{x}^{(t)}_1,\widehat{x}^{(t)}_2,\ldots,\widehat{x}^{(t)}_n)
\end{equation}
match the syndrome by satisfying $\widehat{x}^{(t)} H_1^T= s_x$.
This syndrome match event is called 
\emph{convergence} \cite{poulin2008iterative}.
The BP decoder then outputs $\widehat{E} = D(\widehat{x}^{(t)},0)$ 
as the estimated error operator.
If none of the estimated $\widehat{x}^{(t)}$ match the syndrome 
when $t$ reaches a preset maximum iteration number $T$, 
then BP reports failure due to \emph{non-convergence}.

The computational complexity of BP is $O(nT)$ without parallelism.
If we assume that BP converges 
exponentially fast \cite{dembo2010ising}, 
a natural choice of $T$ would be $O(\log n)$, 
which gives us $O(n\log n)$ for the complexity of BP.

\subsection{Prior Improvements on BP}
\label{subsec:BPOSD_and_BPSI}
BP decoding for QLDPC codes has been known to have issues 
with convergence \cite{poulin2008iterative,raveendran2021trapping}.
Take the $[[882,24,18\le d\le 24]]$ generalized bicycle B1 code
proposed in \cite{Panteleev2021degeneratequantum} for example. 
We simulated its BP decoding performance 
with the sum-product algorithm with $T=100$ 
for the bit-flip noise with error probability $p_x$.
The performance is shown in the blue curve in 
Figure~\ref{fig:B1_BPGD}.
In our simulation, we observe that almost 
all the block error cases are due to non-convergence.

To mitigate the non-convergence issue, Pantaleev and Kalachev 
\cite{Panteleev2021degeneratequantum} proposed to use 
ordered statistics decoding (OSD) as a post-processor 
when BP fails to converge.
In Figure~\ref{fig:B1_BPGD}, 
the red curve shows the BP-OSD performance with order 0 
for the same B1 code over the bit-flip noise. 
In this simulation, 
we employ the normalized min-sum algorithm for BP 
with the normalization factor $\alpha = 0.625$, 
matching the decoder in \cite{Panteleev2021degeneratequantum}.
The BP-OSD decoder shows a significant 
performance improvement over BP. 
This gain has also been observed across various families of quantum 
stabilizer codes \cite{Panteleev2021degeneratequantum,roffe2020decoding}.
However, it is important to note that OSD needs to solve 
a system of linear equations, which costs a 
higher computational complexity of $O(n^3)$ 
\cite{Panteleev2021degeneratequantum} compared to BP. 
In this work, we take BP-OSD as the primary benchmark 
for evaluating the performance of our proposed decoder.

Another improvement on BP referred to as BP 
with stabilizer inactivation (BP-SI)
is proposed by Crest, Mhalla, and Savin in \cite{du2022stabilizer}. 
It starts by sorting the $\pz$-stabilizer rows in $H_1$ 
in increasing order of reliability based on BP soft information. 
After that, the BP algorithm is run $\lambda$ times with exactly 
one of the $\lambda$ least reliable stabilizers inactivated 
(i.e. punctured), with early termination upon convergence.
This BP-SI approach is based on the intuition that 
``stabilizer-splitting'' errors 
\cite{du2022stabilizer, raveendran2021trapping} 
prevent BP convergence. 
In Figure~\ref{fig:B1_BPGD}, the performance of BP-SI with 
$\lambda=10$ on the B1 code is shown in the yellow curve. 
The data points for the yellow curve are taken and translated 
directly from \cite[Figure 2]{du2022stabilizer}, 
where the decoder uses the serial message-passing scheduling 
for the normalized min-sum algorithm with 
normalization factor $\alpha = 0.9$. 
In \cite{du2022stabilizer}, the worst-case complexity and the 
average complexity of BP-SI are claimed to be 
$O(\lambda_{\max}n\log n)$ and $O(\lambda_{\text{avg}}n\log n)$, 
respectively, assuming BP has complexity $O(n\log n)$.
Here, $\lambda_{\max}$ represents the maximum number of 
inactivated stabilizers and $\lambda_{\text{avg}}$ represents 
the average number of inactivated stabilizers.
Notably, at the end of BP-SI, one needs to solve a system 
of linear equations to recover the inactivated error positions.

\section{Belief Propagation with Guided Decimation}
\label{sec:BPGD}

In this section, we introduce a sampling approach
for the syndrome decoding problem of QLDPC codes, 
and show how we can approximate it by 
combining BP with guided decimation (BPGD).
First, we describe the sampling approach 
and show that in the low error rate regime, 
the sampling decoder can achieve performance 
close to the optimal DQML decoder.
Then, we explain how BPGD can be used to 
approximate the sampling decoder. 
Following that, we show the simulated performance of BPGD 
on some QLDPC codes compared against 
BP-OSD and BP-SI under bit-flip noise, 
and analyze its complexity.

\subsection{Sampling Decoding for Stabilizer Codes}
\label{subsec:sampling}

Consider the noise model described in 
Section \ref{subsec:syndrome_decoding}, 
where the encoded qubits of a stabilizer code 
are affected by an $n$-qubit Pauli 
error $E$ following some distribution. 
We now describe a randomized version of QML decoding.
Instead of selecting $\widehat{E}$ with maximal 
posterior probability as in \eqref{eq:qmld}, 
this decoder draws an error $\widehat{E}$ 
from a probability distribution conditioned on 
the observed syndrome event $S=s$. 
Specifically, the probability of an error $D(x,z)$ 
in this conditional probability distribution is given by:
\begin{equation}
\frac{\openone_{(x,z)\Lambda H = s}\cdot
\Pr \big((X,Z)=(x,z))}
{\sum_{(x',z'):(x',z')\Lambda H = s}\Pr \big((X,Z)=(x',z'))},
\label{eq:pmf}
\end{equation}

We note that, following this sampling decoding approach,
each coset $E\mathcal{S}$ of the stabilizer group 
$\mathcal{S}$ in $\mathcal{P}_n$ will also be 
drawn according to their conditional probability 
distribution given syndrome $s$. 
This directly follows from the fact that the 
conditional probability of a coset $E\mathcal{S}$ 
equals the sum of the conditional probabilities for 
the errors in this coset.
Therefore, the sampling version of the DQML decoder 
can be achieved by the sampling version of the QML decoder.
Henceforth, we simply refer to it as the \emph{sampling decoder}.

We remark that the sampling decoder has also been used 
as a theoretical proof technique in works 
on network information theory \cite{yassaee2013technique} 
and classical coding theory \cite{kudekar2016comparing}. 
In the latter, they proved a lemma 
\cite[Lemma 3]{kudekar2016comparing}
that upper bounds the error probability of 
the sampling decoder for a classical code 
by twice the error probability of the
maximum-likelihood (ML) decoder.
A similar upper bound holds for stabilizer codes 
as stated in the following theorem.
The proof of this theorem is deferred to the Appendix.

\begin{theorem}[DQML vs.~sampling]
    \label{thm:DQML_vs_sampling}
    Consider decoding a stabilizer code over 
    Pauli errors following some distribution. 
    Denote the error probability of the DQML decoder by 
    $P_{\mathrm{DQML}}$, 
    and denote the error probability of 
    the sampling decoder by $P_{\mathrm{S}}$.
    Then, the following inequalities hold:
    \begin{equation}
        P_{\mathrm{DQML}} \le P_{\mathrm{S}}
        \le 2\cdot P_{\mathrm{DQML}} 
    \end{equation}
\end{theorem}

Therefore, in the low error rate regime, the sampling decoder can achieve close-to-optimal performance. Next, we describe how we can approximate the sampling decoder with the help of BP.

\subsection{Approximating the Sampling Decoder}
\label{subsec:approx_sampling}

Consider decoding an $[[n,k]]$ CSS code over the bit-flip noise. 
Given the syndrome event $S=s$, the sampling decoder 
would draw a random error vector following a
conditional probability distribution. 
Here, we use the random vector 
\begin{equation}
\widehat{X} = 
(\widehat{X}_1,\widehat{X}_2,\ldots,\widehat{X}_n)
\in\mathbb{F}_2^n
\end{equation}
to denote the output of the sampling decoder.

This sampling process can be achieved by sequentially sampling 
the error bits in $\widehat{X}$ as follows. 
First, we sample $\widehat{X}_1$ according to 
its marginal probabilities
\begin{equation}
\Pr(\widehat{X}_1 = \widehat{x}_1 | S = s),\quad
\widehat{x}_1\in\{0,1\}.
\label{eq:x1_marginal}
\end{equation}
Then, for $i$ from $2$ to $n$, we sequentially sample 
$\widehat{X}_i$ conditioned on both the syndrome $s$ and the 
value of its previous bits following the marginals
\begin{equation}
\Pr(\widehat{X}_i = \widehat{x}_i \,|\, 
\widehat{X}^{i-1}_1=\widehat{x}_1^{i-1}, S = s),
\quad\widehat{x}_i\in\{0,1\}.
\label{eq:xi_marginal}
\end{equation}
Here, we use $\widehat{X}^{i-1}_1$ and $\widehat{x}_1^{i-1}$ 
to denote the subvectors of $\widehat{X}$ and $\widehat{x}$ 
with entries from 1 to $i-1$, respectively.

The error sampled this way has the correct conditional 
probability according to the chain rule:
\begin{multline}
\Pr(\widehat{X} = \widehat{x} \,|\, S = s) = \\
\prod_{i=1}^n 
\Pr(\widehat{X}_i = \widehat{x}_i \,|\, 
\widehat{X}^{i-1}_1=\widehat{x}_1^{i-1}, S = s).
\end{multline}
Note that we are also free to choose an arbitrary order 
to sample the bits in $\widehat{X}$. 

Since BP can be used to compute the 
approximated marginal probabilities for the error bits 
as shown in \eqref{eq:marginal}, 
the marginals in \eqref{eq:x1_marginal} and \eqref{eq:xi_marginal} 
can be estimated by running BP on a Tanner graph.
Therefore, we can approximate the above sampling process by
iterating the following steps:
\begin{enumerate}[1)]
    \item Run BP to obtain estimated marginals for the 
    remaining bits in the error vector.
    \item Choose an $\widehat{X}_i$ and fix its value to 
    $\widehat{x}_i$ according to its estimated marginals.
    \item Update the condition to include 
    $\widehat{X}_i = \widehat{x}_i$.
\end{enumerate}
The process of sequentially fixing variables to hard decisions during iterative decoding has been referred to as guided decimation ~\cite{mezard2002analytic,montanari2007solving}. 
We note that in \cite{montanari2007solving}, 
BP guided decimation combined with warning propagation can also be understood to approximate the process of sampling a vector from the distribution 
implied by the Tanner graph for constraint satisfaction problems.
\subsection{The BPGD Algorithm}
\label{subsec:BPGD_algo}

Message-passing algorithms with ``decimation'' were first introduced for the $K$-SAT constraint satisfaction problem based on insights from statistical physics~\cite{mezard2002analytic}.
In such problems, there are typically many valid solutions and the goal is to find just one of them.
In~\cite{mezard2002analytic}, decimation was first combined with a related message-passing algorithm called survey propagation.
Later, the idea of decimation was extended to define the BP guided decimation (BPGD) algorithm~\cite{montanari2007solving} and related approaches were applied to the lossy compression problem~\cite{wainwright2005lossy,filler2007binary,aref2015approaching}.

Now, following the setup in Section~\ref{subsec:syndrome_bp},
we describe the BPGD algorithm for decoding bit-flip errors 
in detail on a $[[n,k]]$ CSS code.

\begin{algorithm}[t]
\caption{BPGD over bit flips}
\label{alg:BPGD}
\KwIn{
block length $n$, \\
Tanner graph $G = (V,C,E)$ for $H_1$,\\
syndrome $s_x$, \\
bit-flip error rate $p_x$, \\
number of BP iterations per round $T$}
\KwOut{estimated $\widehat{x}$ or non-convergence}
$\mu_{v_i} = \log((1-p_x)/p_x)$ for all $v_i\in V$\\
$m^{(0)}_{v_i\rightarrow c_j} = \mu_{v_i}$ for all 
$v_i\in V,c_j\in\partial v_i$ \\
$V_u = V$ \\
\For{$r=1$ \emph{\KwTo} $n$}{
    run BP for $T$ iterations\\ 
    $\widehat{x} \leftarrow$ hard values of the variable nodes\\
    \eIf{$\widehat{x}H_1^T = s_x$}{
        \Return{$\widehat{x}$}
    }{
        $v_i = \arg\max_{v\in V_u}\gamma(v_i)$ \\
        \eIf{$m_{v_i}^{(rT)} \geq 0$}{
            $\mu_{v_i} = \text{llr}_{\max}$
        }{
            $\mu_{v_i} = -\text{llr}_{\max}$
        }
        $V_u = V_u\backslash\{v_i\}$
    }
}
\Return{non-convergence}
\end{algorithm}

First, we initialize the channel LLR for each variable node 
$v_i$ on the Tanner graph as
\begin{equation}
\mu_{v_i}=
\log\left(\frac{1-p_x}{p_x}\right)
\end{equation}
similar to BP.

Then, BPGD proceeds in rounds with $r$ going from 1 to $n$.
In the $r$-th round, we first run BP with the sum-product algorithm 
(\eqref{eq:check_node_update} and \eqref{eq:bit_node_update}) 
on the Tanner graph for $T$ iterations to obtain the estimated 
marginals for the variable nodes.
If BP converges to an error matching the syndrome, 
then the decoder immediately terminates and 
outputs the hard values of the variable nodes as the estimated error.

Otherwise, out of all the variable nodes that are not yet decimated,
we pick $v_i$ with the largest reliability $\gamma(v_i)$ 
in \eqref{eq:gamma} for decimation.
We decimate this variable node $v_i$ by updating its channel message 
$\mu_{v_i}$ based on its bias $m_{v_i}^{(rT)}$ as
\begin{equation}
\mu_{v_i} = 
\begin{cases}
\text{llr}_{\max}, & \text{if } m_{v_i}^{(rT)} > 0 \\
-\text{llr}_{\max}, & \text{if } m_{v_i}^{(rT)} \leq 0,
\end{cases}
\label{eq:llr_update}
\end{equation}
where $\text{llr}_{\max}$ is a large fixed number. 

If we choose $\text{llr}_{\max}$ to be infinity 
(i.e. $\text{llr}_{\max} = \infty$), 
this decimation effectively assigns a hard value for 
$\widehat{X}_i$ as $\widehat{x}_i$ according to the bias of $v_i$. 
Different from the sampling decoder that samples $\widehat{X}_i$ 
according to its marginals, 
here we simply fix the value of $\widehat{X}_i$ towards its bias.
The new condition $\widehat{X}_i = \widehat{x}_i$
is also incorporated into the Tanner graph 
and takes effect during BP in the subsequent rounds.
While letting $\text{llr}_{\max}$ to be infinity 
makes sense in theory, 
it can introduce numerical issues in software implementation.
and thus setting $\text{llr}_{\max}$ to be a large fixed number 
is preferred in practice.
For all our simulation results shown in this work, 
we set $\text{llr}_{\max} = 25$. 

After decimating the most reliable variable node, 
we proceed to the next round $r+1$ by continuing to run 
BP for another $T$ iterations followed by decimation.
This process continues until either BP converges 
or all the variable nodes have been decimated.
If, after all variable nodes are decimated, their hard values 
still do not match the syndrome, then this is referred to as a 
\emph{non-convergence failure} of the BPGD algorithm. 
The pseudo-code for the BPGD algorithm is provided in 
Algorithm \ref{alg:BPGD}.

\subsection{Numerical Results}
In Section \ref{subsec:BPGD_algo}, we describe BPGD as an approximation for the sampling decoder, where we sequentially decimate variable nodes after each round of BP. 
Here we show simulation results of BPGD comparing two different decimation orders: sequentially decimating the most reliable variable nodes as described in Algorithm \ref{alg:BPGD}, versus randomly decimating variable nodes.
For the second decimation order, after each round of BP, we randomly pick a variable node that hasn't been decimated and update its channel message according to its bias as in \eqref{eq:llr_update}.

\begin{figure}[t]
\centering
\begin{tikzpicture}
\begin{axis}[
scale=0.95,
legend cell align={left},
legend style={
  fill opacity=0.7,
  draw opacity=1,
  text opacity=1,
  at={(1,0)},
  anchor=south east,
  draw=lightgray204,
  font=\scriptsize
},
grid=both,
label style = {font=\footnotesize},
ticklabel style = {font=\footnotesize},
grid style={darkgray176},
xmin=0.035, xmax=0.105,
xmajorgrids,
xminorgrids,
xtick={0.04,0.05,0.06,0.07,0.08,0.09,0.10},
xticklabels={0.04,0.05,0.06,0.07,0.08,0.09,0.10},
xlabel={$p_x$},
ymode=log,
ymin=5e-7, ymax=2.0,
yminorgrids,
ylabel={Block Error Rate},
tick pos=left,
]

\addplot [thick, color1, mark=*]
table {%
0.04 0.043400 
0.05 0.086700
0.06 0.144900
0.07 0.250500
0.08 0.444800
0.09 0.714100
0.10 0.913900
};
\addlegendentry{BP}

\addplot [thick, color6, mark=*]
table {
0.04 0.000023
0.05 0.000365
0.06 0.005857
0.07 0.061631
0.08 0.268817
0.09 0.578035
0.10 0.862069
};
\addlegendentry{BPGD, $T=100$, rd-order}

\addplot [thick, color5, mark=*]
table {%
0.04 0.000001
0.05 0.000095
0.06 0.003031
0.07 0.036055
0.08 0.247002
0.09 0.517766
0.10 0.814516
};
\addlegendentry{BPGD, $T=100$}

\end{axis}
\end{tikzpicture}
\caption{Performance of BPGD on the $[[882,24,18\le d\le 24]]$ B1 code \cite{Panteleev2021degeneratequantum} over bit-flip noise following different decimation orders. 
The data points are collected by running simulations until we observe 100 error cases.}
\label{fig:B1_BPGD_dec_order}
\end{figure}
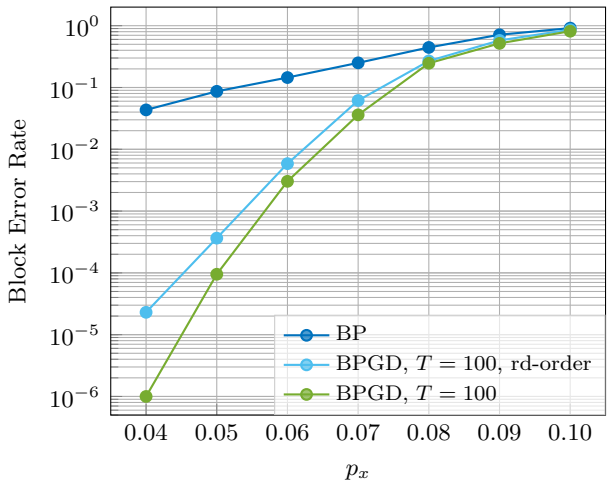

In Figure~\ref{fig:B1_BPGD_dec_order}, we show the simulation result of BPGD for the $[[882,24,18\le d\le 24]]$ B1 code \cite{Panteleev2021degeneratequantum} over bit-flip noise with error probability $p_x$.
We start by setting $T=100$ to ensure that in each round, 
BP is run for a sufficient number of iterations to provide 
accurate approximate marginal probabilities for the variable nodes.
First, we observe that BPGD provides a significant performance gain compared with ordinary BP.
Then, compared with random decimation order as shown in the light blue curve, BPGD which sequentially decimates the most reliable variable nodes offers better performance, as shown in the green curve.

\begin{figure}[t]
\centering
\begin{tikzpicture}
\begin{axis}[
scale=0.95,
legend cell align={left},
legend style={
  fill opacity=0.7,
  draw opacity=1,
  text opacity=1,
  at={(1,0)},
  anchor=south east,
  draw=lightgray204,
  font=\scriptsize
},
grid=both,
label style = {font=\footnotesize},
ticklabel style = {font=\footnotesize},
grid style={darkgray176},
xmin=0.035, xmax=0.105,
xmajorgrids,
xminorgrids,
xtick={0.04,0.05,0.06,0.07,0.08,0.09,0.10},
xticklabels={0.04,0.05,0.06,0.07,0.08,0.09,0.10},
xlabel={$p_x$},
ymode=log,
ymin=5e-7, ymax=2.0,
yminorgrids,
ylabel={Block Error Rate},
tick pos=left,
]

\addplot [thick, color1, mark=*]
table {%
0.04 0.043400
0.05 0.086700
0.06 0.144900
0.07 0.250500
0.08 0.444800
0.09 0.714100
0.10 0.913900
};
\addlegendentry{BP}

\addplot [thick, color2, mark=*]
table {
0.04 0.000029
0.05 0.000391
0.06 0.006060
0.07 0.053763
0.08 0.248768
0.09 0.534759
0.10 0.855932
};
\addlegendentry{BPGD, $T=1$}

\addplot [thick, color4, mark=*]
table {
0.04 0.000006
0.05 0.000179
0.06 0.003891
0.07 0.038801
0.08 0.216942
0.09 0.595376
0.10 0.863248
};
\addlegendentry{BPGD, $T=10$}

\addplot [thick, color6, mark=*]
table {
0.04 0.000002
0.05 0.000116
0.06 0.003703
0.07 0.042721
0.08 0.213836
0.09 0.507538
0.10 0.834711
};
\addlegendentry{BPGD, $T=70$}

\addplot [thick, color5, mark=*]
table {
0.04 0.000001
0.05 0.000095
0.06 0.003031
0.07 0.036055
0.08 0.247002
0.09 0.517766
0.10 0.814516
};
\addlegendentry{BPGD, $T=100$}

\end{axis}
\end{tikzpicture}
\caption{Performance of BPGD on the $[[882,24,18\le d\le 24]]$ B1 code \cite{Panteleev2021degeneratequantum} over bit-flip noise with different BP iterations per round. 
The data points are collected by running simulations until we observe 100 error cases.}
\label{fig:B1_BPGD_T}
\end{figure}
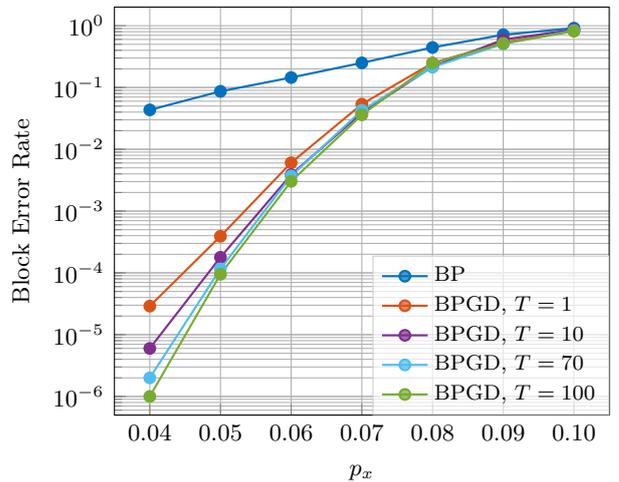

In Figure \ref{fig:B1_BPGD_T}, we also compare the BPGD performance with different BP iterations per round. We see that while $T=100$ provides the best performance, reducing the number of BP iterations per round from $T=100$ to $T=10$, shown by the purple curve, 
does not significantly degrade the BPGD performance. 

\subsection{Performance Comparison}

In Figure \ref{fig:B1_BPGD}, we compare our BPGD performance with some prior works for the $[[882,24,18\le d\le 24]]$ B1 code 
\cite{Panteleev2021degeneratequantum} over bit-flip noise.
The performance of BPGD with $T=10$, shown by the purple curve, 
is identical to the purple curve in Figure~\ref{fig:B1_BPGD_T}. 
For comparison, Figure \ref{fig:B1_BPGD} includes the decoding performance of BP, BP-OSD with order 0, and BP-SI with $\lambda=10$, whose settings are given in Section~\ref{subsec:BPOSD_and_BPSI}.
BPGD with $T=10$ yields the best performance. 
It is worth noting that, in our simulations, 
the majority of the errors (over 90\%) from the BPGD runs 
for both $T=100$ and $T=10$ are attributed to non-convergence. 
Therefore, similar to the BP algorithm, we rarely encounter logical errors upon convergence from running BPGD.

\begin{figure}[t]
\centering
\begin{tikzpicture}
\begin{axis}[
scale=0.95,
legend cell align={left},
legend style={
  fill opacity=0.7,
  draw opacity=1,
  text opacity=1,
  at={(1,0)},
  anchor=south east,
  draw=lightgray204,
  font=\scriptsize
},
grid=both,
label style = {font=\footnotesize},
ticklabel style = {font=\footnotesize},
grid style={darkgray176},
xmin=0.035, xmax=0.105,
xmajorgrids,
xminorgrids,
xtick={0.04,0.05,0.06,0.07,0.08,0.09,0.10},
xticklabels={0.04,0.05,0.06,0.07,0.08,0.09,0.10},
xlabel={$p_x$},
ymode=log,
ymin=1e-6, ymax=2.0,
yminorgrids,
ylabel={Block Error Rate},
tick pos=left,
]

\addplot [thick, color1, mark=*]
table {%
0.04 0.043400 
0.05 0.086700
0.06 0.144900
0.07 0.250500
0.08 0.444800
0.09 0.714100
0.10 0.913900
};
\addlegendentry{BP}

\addplot [thick, color2, mark=*]
table {%
0.04 0.000020
0.05 0.000930
0.06 0.016510
0.07 0.119910
0.08 0.404500
0.09 0.732800
0.10 0.920000
};
\addlegendentry{BP-OSD-0}

\addplot [thick, color3, mark=*]
table {%
0.06667 0.026000
0.06000 0.005500
0.05333 0.000700
0.04667 0.000070
0.04000 0.000008
};
\addlegendentry{BP-SI, $\lambda=10$ \cite{du2022stabilizer}} 

\addplot [thick, color4, mark=*]
table {%
0.04 0.00000321835
0.05 0.00016846279
0.06 0.00367352876
0.07 0.04541446208
0.08 0.24343675417
0.09 0.54594594594
0.10 0.87179487179
};
\addlegendentry{BPGD, $T=10$}

\end{axis}
\end{tikzpicture}
\caption{Performance of various decoders on the $[[882,24,18\le d\le 24]]$ B1 code \cite{Panteleev2021degeneratequantum} over bit-flip noise. 
The data points of BP, BP-OSD with order 0, and BPGD decoders are collected by running simulations until we observe 100 error cases.
The data points of BP-SI are taken and translated from \cite[Figure 2]{du2022stabilizer}}
\label{fig:B1_BPGD}
\end{figure}

\begin{figure}[t]
\centering
\begin{tikzpicture}
\begin{axis}[
scale=0.95,
legend cell align={left},
legend style={
  fill opacity=0.7,
  draw opacity=1,
  text opacity=1,
  at={(1,0)},
  anchor=south east,
  draw=lightgray204,
  font=\scriptsize
},
grid=both,
label style = {font=\footnotesize},
ticklabel style = {font=\footnotesize},
grid style={darkgray176},
xmin=0.035, xmax=0.105,
xmajorgrids,
xminorgrids,
xtick={0.04,0.05,0.06,0.07,0.08,0.09,0.10},
xticklabels={0.04,0.05,0.06,0.07,0.08,0.09,0.10},
xlabel={$p_x$},
ymode=log,
ymin=5e-5, ymax=2.0,
yminorgrids,
ylabel={Block Error Rate},
tick pos=left,
]

\addplot [thick, color1, mark=*]
table {%
0.04 0.080058
0.05 0.157581
0.06 0.268469
0.07 0.400398
0.08 0.636991
0.09 0.872285
0.10 0.983284
};
\addlegendentry{BP}

\addplot [thick, color2, mark=*]
table {%
0.10 0.993000
0.09 0.913000
0.08 0.588000
0.07 0.166400
0.06 0.028600
0.05 0.005100
0.04 0.000700
};
\addlegendentry{BP-OSD-0}

\addplot [thick, color3, mark=*]
table {%
0.06667 0.025000
0.06000 0.008000
0.05333 0.002000
0.04667 0.000600
0.04000 0.000180
};
\addlegendentry{BP-SI, $\lambda=10$ \cite{du2022stabilizer}} 

\addplot [thick, color4, mark=*]
table {%
0.10 0.925000
0.09 0.628000
0.08 0.209000
0.07 0.035000
0.06 0.005800
0.05 0.000850
0.04 0.000144
};
\addlegendentry{BPGD, $T=10$}
\end{axis}
\end{tikzpicture}
\caption{Performance of various decoders on the $[[1922,50,16]]$ C2 code \cite{Panteleev2021degeneratequantum} over bit-flip noise.
The data points of BP, BP-OSD with order 0, and BPGD decoders are collected by running simulations until we observe 100 error cases.
The data points of BP-SI are taken and translated from \cite[Figure 2]{du2022stabilizer}}
\label{fig:C2_BPGD}
\end{figure}
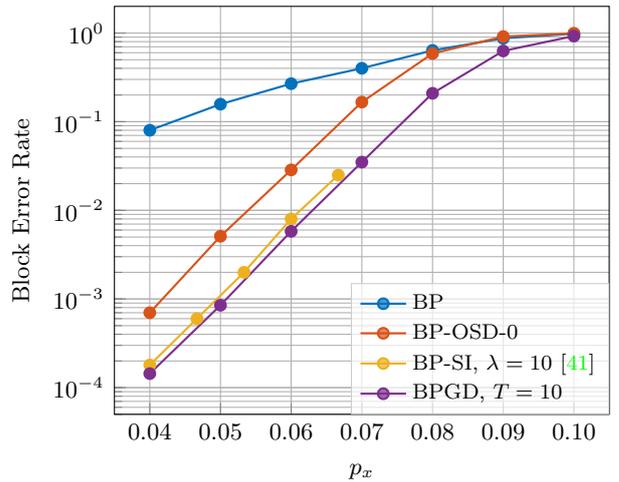

In Figure \ref{fig:C2_BPGD}, we also show the simulation result 
of BPGD with $T=10$ for the $[[1922,50,16]]$ hypergraph product 
code C2 \cite{Panteleev2021degeneratequantum} over bit-flip noise 
with error probability $p_x$. 
For comparison, we include the performances for BP, 
BP-OSD with order 0, and BP-SI with $\lambda=10$, 
whose respective settings remain consistent with those 
in Figure \ref{fig:B1_BPGD}:
1) the BP decoder runs the basic sum-product algorithm with 
$T=100$; 
2) the BP-OSD decoder with order 0 runs the serial 
normalized min-sum algorithm with 
normalization factor $\alpha=0.625$;
3) the BP-SI decoder with $\lambda=10$ runs the serial 
normalized min-sum algorithm with 
normalization factor $\alpha=0.9$, 
whose data points are taken and translated 
directly from \cite[Figure 2]{du2022stabilizer}.
We can see that for the two QLDPC codes we considered in Figure~\ref{fig:B1_BPGD} and Figure~\ref{fig:C2_BPGD}, 
BPGD shows better performance compared with 
BP-OSD with order 0 and BP-SI with $\lambda = 10$.

\subsection{Complexity Analysis for BPGD}
BPGD requires at most $n$ decimation rounds, each involving $T$ iterations of message-passing updates of complexity $O(n)$ and
a search for the most reliable qubit for decimation with complexity $O(n)$. 
Therefore, the worst-case complexity of BPGD is $O(T n^2)$. 
This worst-case complexity scales roughly the same as BP-SI \cite{du2022stabilizer}, and it is better compared with BP-OSD with order 0, which has complexity $O(n^3)$ \cite{Panteleev2021degeneratequantum}. 
Moreover, unlike BP-OSD and BP-SI, BPGD does not require solving systems of linear equations, which makes it potentially more friendly for hardware implementation.

If we assume BP in each round has complexity $O(Tn)$, then the average-case complexity of BPGD becomes $O(r_{\text{avg}}Tn)$, where $r_{\text{avg}}$ denotes the average number of decimated variable nodes.
The value of $r_{\text{avg}}$ is highly dependent on the bit-flip error probability $p_x$, meaning BPGD has different average complexity in different error rate regimes. 
Table~\ref{tab:BPGD_ave_rds} shows the average number of decimated variable nodes for BPGD with $T = 10$ when decoding the B1 code.
Note that when calculating $r_{\text{avg}}$ in Table \ref{tab:BPGD_ave_rds}, we take into account both the convergent cases and the non-convergent cases. 
In the non-convergence cases, the number of variable nodes decimated by BPGD equals the block length $n=882$. 
From Table \ref{tab:BPGD_ave_rds} we can see that, in the low error rate regime such as $p_x = 0.05$, $r_{\text{avg}}$ becomes very small, making the average complexity of BPGD approaches the BP complexity $O(Tn)$. 
We note that a similar observation has also been made for BP-SI concerning the average number of inactivated stabilizers in the low error rate regime \cite{du2022stabilizer}.

\begin{table}[t]
    \centering
    \scalebox{0.95}{
    \begin{tabular}{|c|c|c|c|c|}
        \hline
        $p_x$ & 0.05 & 0.06 & 0.07 & 0.08 \\\hline
        sim. runs & 1000000 & 100000 & 100000 & 10000 \\\hline
        $r_{\text{avg}}$ & 2.91 & 9.82 & 60.46 & 231.7 \\
        \hline
    \end{tabular}
    }
    \caption{Average number of decimation rounds of BPGD with $T=100$ on the B1 code \cite{Panteleev2021degeneratequantum} over bit-flip noise.}
    \label{tab:BPGD_ave_rds}
\end{table}

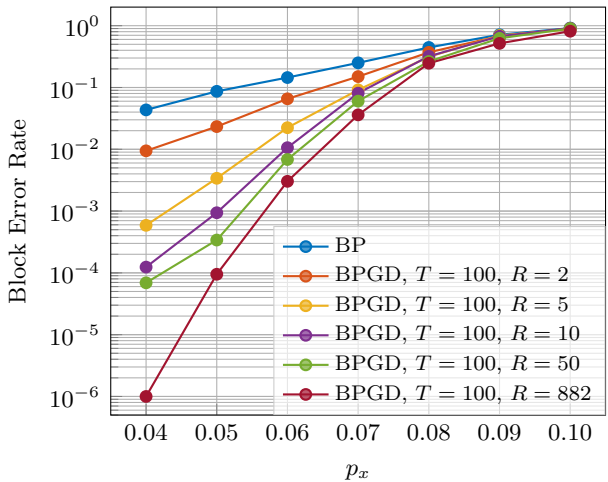
\begin{figure}[t]
\centering
\begin{tikzpicture}
\begin{axis}[
scale=0.95,
legend cell align={left},
legend style={
  fill opacity=0.7,
  draw opacity=1,
  text opacity=1,
  at={(1,0)},
  anchor=south east,
  draw=lightgray204,
  font=\scriptsize
},
grid=both,
label style = {font=\footnotesize},
ticklabel style = {font=\footnotesize},
grid style={darkgray176},
xmin=0.035, xmax=0.105,
xmajorgrids,
xminorgrids,
xtick={0.04,0.05,0.06,0.07,0.08,0.09,0.10},
xticklabels={0.04,0.05,0.06,0.07,0.08,0.09,0.10},
xlabel={$p_x$},
ymode=log,
ymin=5e-7, ymax=2.0,
yminorgrids,
ylabel={Block Error Rate},
tick pos=left,
]

\addplot [thick, color1, mark=*]
table {%
0.04 0.043400 
0.05 0.086700
0.06 0.144900
0.07 0.250500
0.08 0.444800
0.09 0.714100
0.10 0.913900
};
\addlegendentry{BP}

\addplot [thick, color2, mark=*]
table {%
0.04 0.009440
0.05 0.023300
0.06 0.065600
0.07 0.150100
0.08 0.370800
0.09 0.680100
0.10 0.904800
};
\addlegendentry{BPGD, $T=100$, $R = 2$}

\addplot [thick, color3, mark=*]
table {%
0.04 0.000588
0.05 0.003396
0.06 0.022356
0.07 0.091491
0.08 0.313480
0.09 0.694444
0.10 0.877193
};
\addlegendentry{BPGD, $T=100$, $R = 5$}

\addplot [thick, color4, mark=*]
table {%
0.04 0.000124
0.05 0.000940
0.06 0.010620
0.07 0.081000
0.08 0.320000
0.09 0.675000
0.10 0.910000
};
\addlegendentry{BPGD, $T=100$, $R = 10$}

\addplot [thick, color5, mark=*]
table {%
0.04 0.000069
0.05 0.000340
0.06 0.006838
0.07 0.060241
0.08 0.263852
0.09 0.625000
0.10 0.900901
};
\addlegendentry{BPGD, $T=100$, $R = 50$}

\addplot [thick, color7, mark=*]
table {%
0.04 0.000001
0.05 0.000095
0.06 0.003031
0.07 0.036055
0.08 0.247002
0.09 0.517766
0.10 0.814516
};
\addlegendentry{BPGD, $T=100$, $R = 882$}

\end{axis}
\end{tikzpicture}
\caption{Performance of BPGD with different decimation round limits on the $[[882,24,18\le d\le 24]]$ B1 code \cite{Panteleev2021degeneratequantum} over bit-flip noise.}
\label{fig:B1_BPGD_rd_lim}
\end{figure}

One natural way to reduce the worst-case complexity of BPGD is by bounding the maximum number of variable nodes it decimates. 
This can be achieved by modifying line 4 in Algorithm~\ref{alg:BPGD} to ``for $r = 1$ to $R$ do'', where $R <n$ represents a predefined limit on the number of decimation rounds.
With this modification, the worst-case complexity of BPGD becomes $O(RTn)$. 
In Figure \ref{fig:B1_BPGD_rd_lim}, we present the decoding performance of BPGD with $T=100$ and different round limits $R$ for the $[[882,24,18\le d\le 24]]$ B1 code \cite{Panteleev2021degeneratequantum}. 
As expected, the performance of BPGD steadily improves with increasing round limit $R$. 
The optimal performance is attained when $R$ equals the block length $n=882$. 
This allows us to pick a desired trade-off between worst-case complexity and the BPGD decoding performance tailored to specific application requirements.

\section{Randomized BPGD: An Experiment}
\label{sec:BPGD_rd}
For the syndrome decoding of degenerate QLDPC codes, one reason for the non-convergence issue of BP is that there may exist multiple low-weight error patterns that match the syndrome \cite{poulin2008iterative}. 
Due to code degeneracy, many of them differ only by a stabilizer. 
Intuitively, this is supposed to help the decoder, as it yields the same decoding result by picking any of those solutions. 
However, when running BP, the locally operating algorithm gets confused about the direction to proceed, resulting in non-convergence \cite{raveendran2021trapping}.
In a case study presented in \cite[Section IV. A]{poulin2008iterative}, Poulin and Chung explored this scenario for a two-qubit stabilizer code.
A more comprehensive study of this phenomenon from the perspective of trapping sets can be found in the work by Raveendran and Vasi\'c \cite{raveendran2021trapping}. 
In another work, Kuo and Lai \cite{kuo2022exploiting} examined this situation by regarding BP as an energy-minimization process, and explain non-convergence as being trapped in a local minimum of the energy function.

Here, we devise an experiment to shed light on this situation. 
Consider the following randomized version of the BPGD algorithm, where in each round, instead of decimating the most reliable variable node, we identify the top few reliable variable nodes and then randomly select one for decimation.
Concretely, this randomized decimation process goes as follows.
Denote $V_u$ as the set of variable nodes that have not yet been decimated.
In the $r$-th round, after running BP for $T$ iterations, we first find the most reliable variable node in $V_u$ and denote its reliability as $\gamma_{\max}$. 
Then, we construct a set $P$ including all $v_i\in V_u$ whose reliabilities satisfy $\gamma(v_i) \ge \gamma_{\max} - \gamma'$.
Here, $P$ is a set of many reliable variable nodes prepared for random decimation, and $\gamma'$ stands for a gap such that the reliability threshold for $P$ is $\gamma_{\max} - \gamma'$.
Note that $P$ has size at least one 
since it always includes the most reliable variable node in $V_u$.
After that, we randomly select a variable node in $P$ for decimation.
The pseudo-code for the randomized BPGD algorithm, referred to as BPGD-rd, is presented in Algorithm~\ref{alg:BPGD_rd}.
The differences between BPGD-rd and BPGD in 
Algorithm~\ref{alg:BPGD} regarding the randomized decimation process are shown in lines 10-12 in Algorithm~\ref{alg:BPGD_rd}.

\begin{algorithm}[t]
\caption{BPGD-rd over bit flips}
\label{alg:BPGD_rd}
\KwIn{
block length $n$, \\
Tanner graph $G = (V,C,E)$ for $H_1$,\\
syndrome $s_x$, \\
bit-flip error rate $p_x$, \\
number of BP iterations per round $T$}
\KwOut{estimated $\widehat{x}$ or non-convergence}
$\mu_{v_i} = \log((1-p_x)/p_x)$ for all $v_i\in V$\\
$m^{(0)}_{v_i\rightarrow c_j} = \mu_{v_i}$ for all 
$v_i\in V,c_j\in\partial v_i$ \\
$V_u = V$ \\
\For{$r=1$ \emph{\KwTo} $n$}{
    run BP for $T$ iterations\\ 
    $\widehat{x} \leftarrow$ hard values of the variable nodes\\
    \eIf{$\widehat{x}H_1^T = s_x$}{
        \Return{$\widehat{x}$}
    }{
        $\gamma_{\max} = \max_{v_i\in V_u}\gamma(v_i)$ \\
        $P = \{v_i\in V_u \;|\; \gamma(v_i) \ge \gamma_{\max} - \gamma'\}$ \\
        randomly select $v_i$ from $P$ \\
        \eIf{$m_{v_i}^{(rT)} \geq 0$}{
            $\mu_{v_i} = \text{llr}_{\max}$
        }{
            $\mu_{v_i} = -\text{llr}_{\max}$
        }
        $V_u = V_u\backslash\{v_i\}$
    }
}
\Return{non-convergence}
\end{algorithm}

Since we are randomly selecting variable nodes in BPGD-rd each round 
for decimation, this algorithm can potentially converge to many different errors.
Leveraging this property, we consider the following experiment for the $[[882,24,18\le d\le 24]]$ B1 code \cite{Panteleev2021degeneratequantum}.
First, we select a bit-flip error $X$ with weight 73. 
This error is randomly generated with $p_x=0.08$. 
Then, we run the BPGD-rd algorithm 10000 times to decode its syndrome, each with a distinct seed for the random decimation process. 
The configuration for BPGD-rd during this experiment includes $T=10$ for the number of BP iterations per round, $p_x = 0.08$ for the bit-flip error probability, and $\gamma' = 1.0$ for the reliability gap.

We observe that out of 10000 BPGD-rd runs, 9568 of them converge. 
Within these 9568 convergent cases, we identify a total of 111 distinct errors, whose weights range from 73 to 87.
Table \ref{tab:BPGD_rd} shows the 10 most frequent errors along with their weights and distances relative to the channel error $X$.
Notably, all 111 error patterns differ from $X$ by a stabilizer, which means that all of them decode $X$ successfully due to degeneracy.
If we run the original BPGD algorithm in Algorithm \ref{alg:BPGD} 
for decoding, it converges to the error with index 2 in Table \ref{tab:BPGD_rd}.

\begin{table}[t]
    \centering
    \scalebox{0.95}{
    \begin{tabular}{|c|c|c|c|}
    \hline
    index & frequency & weight & distance to $X$\\\hline
    1 & 2944 & 73 & 22 \\\hline
    2 & 2727 & 73 & 26 \\\hline
    3 & 1560 & 73 & 20 \\\hline
    4 & 1521 & 73 & 16 \\\hline
    5 & 125 & 73 & 30 \\\hline
    6 & 110 & 73 & 26 \\\hline
    7 & 91 & 73 & 16 \\\hline
    8 & 68 & 73 & 20 \\\hline
    9 & 66 & 73 & 24 \\\hline
    10 & 32 & 75 & 28 \\\hline
    \end{tabular}
    }
    \caption{A list of 10 most frequent errors 
    obtained out of 10000 BPGD-rd decoding runs
    on B1 code \cite{Panteleev2021degeneratequantum} 
    for a bit-flip error $X$ with weight 73.}
    \label{tab:BPGD_rd}
\end{table}

From this experiment we see that for the syndrome decoding problem of a degenerate QLDPC code such as the B1 code, there can be multiple low-weight error patterns matching the input syndrome, confusing the BP decoding process. 
In particular, for decoding $X$ in this experiment, BP itself with the sum-product algorithm results in non-convergence. 
Moreover, it also shows that combining BP with guided decimation encourages BP convergence towards one of those errors.
In this case, degeneracy plays to our advantage since decoding is successful no matter which of the 111 errors is selected.

Here we make some additional remarks regarding this experiment.
Firstly, it can be observed that the distribution of the error patterns upon convergence in the BPGD-rd algorithm is not uniform.
We can see from Table \ref{tab:BPGD_rd} that, for the syndrome decoding problem we investigated, a significant portion of the convergent cases focuses on four specific weight-73 errors, indexed by 1-4 in Table \ref{tab:BPGD_rd}. 
In contrast, the original weight-73 bit-flip error $X$ does not even appear among the 111 errors we observed.

Secondly, we note that due to the random decimation process inherent to BPGD-rd, the results presented in this section are the product of a single simulation run of this experiment.
A repeat of the same experiment could yield different numerical results. 
Additionally, the outcomes depend on the specific choice of the channel error $E_x$. 
However, we observe that the phenomenon of multiple distinct error patterns matching the input syndrome and the uneven distribution among convergent error patterns appears to be a general characteristic of the BPGD-rd algorithm.

\section{Quaternary Belief Propagation with Guided Decimation}
\label{sec:QBPGD}

In this section, we consider the syndrome decoding problem of a $[[n,k]]$ CSS code over depolarizing noise, and present a natural extension of the binary BPGD algorithm that gives the quaternary version.

In the depolarizing noise model with physical error rate $p$, each encoded qubit is independently affected by a Pauli $\px$, $\py$, or $\pz$ error, each occurring with a probability of $p/3$. 
Here, we represent the Pauli error as a random quaternary vector $Q=(Q_1,\ldots,Q_n)$ with its realization $q = (q_1,\ldots,q_n)\in\{0,1,2,3\}^n$, with $0,1,2,3$ representing the absence of error, or the presence of a Pauli $\px$, $\py$ or $\pz$ error, respectively.
We define the Pauli error represented by $q$ as $E(q)$. 
For example, 
\begin{equation}
E(q=(1,0,2,3,0)) = \px\otimes I\otimes \py\otimes \pz\otimes I.
\end{equation}

For depolarizing noise, we consider the quaternary BP (Q-BP) algorithm 
\cite{poulin2008iterative}, also referred to as the 
non-binary version of the syndrome BP algorithm \cite{babar2015fifteen}.
Notably, Q-BP offers better performance over depolarizing noise in comparison to decoding bit-flip errors and phase-flip errors separately with binary BP, as exemplified in \cite[Figure 5]{Panteleev2021degeneratequantum}.
For detailed descriptions of Q-BP, we refer the readers to \cite{kuo2020refined} and to \cite[Algorithm 1]{babar2015fifteen}. 
An efficient log domain implementation of Q-BP has also been discussed in \cite{lai2021log}.

For the syndrome decoding problem on a CSS code, Q-BP runs on a Tanner graph $G=(V,C,E)$ similar to binary BP, except that here $C$ contains check nodes representing both the $\px$-stabilizers in $H_2$ and the $\pz$-stabilizers in $H_1$, where $H_1$ and $H_2$ are 
submatrices of $H_x$ and $H_z$, respectively, as in (\ref{eq:H1_G2}). 

\begin{algorithm}[t!]
\caption{Q-BPGD for depolarizing noise}
\label{alg:Q-BPGD}
\KwIn{block length $n$, Tanner graph $G$, \\
syndrome $s$, physical error rate $p$, \\
number of iterations per round $T$}
\KwOut{estimated $\widehat{q}$ or non-convergence}
$\mu_{v_i} = (1-p,\;p/3,\;p/3,\;p/3)$ for all $v_i\in V$\\
$V_u = V$ \\
\For{$r=1$ \emph{\KwTo} $n$}{
    run Q-BP for $T$ iterations \\
    $\widehat{q} \leftarrow$ hard values for the variable nodes\\
    \eIf{$\widehat{q}$ matches the syndrome $s$}{
        \Return{$\widehat{q}$}
    }{
        $v_i = \arg\max_{v\in V_u}\gamma(v_i)$ \\
        \Switch{$\underset{j\in\{0,1,2,3\}}{\arg\max}\;p_{v_i,j}$}
        {
            \lCase{$0$}{$\mu_{v_i} = (1-\epsilon,\epsilon,\epsilon,\epsilon)$}
            \lCase{$1$}{$\mu_{v_i} = (\epsilon,1-\epsilon,\epsilon,\epsilon)$}
            \lCase{$2$}{$\mu_{v_i} = (\epsilon,\epsilon,1-\epsilon,\epsilon)$}
            \lCase{$3$}{$\mu_{v_i} = (\epsilon,\epsilon,\epsilon,1-\epsilon)$}
        }
        $V_u = V_u\backslash\{v_i\}$
    }
}
\Return{non-convergence}
\end{algorithm}

In addition, the channel LLRs for the variable nodes are now become the channel message
\begin{equation}
\mu_{v_i} = (1-p,\;p/3,\;p/3,\;p/3), 
\end{equation}
and each variable node $v_i\in V$ contains information of four normalized probabilities 
\begin{equation}
(p_{v_i,0},\;p_{v_i,1},\;p_{v_i,2},\;p_{v_i,3})
\end{equation}
which, after a sufficient number of BP iterations,
approximate the marginal probabilities for 
$Q_i$ conditioned on the syndrome $s$ as 
\begin{equation}
    p_{v_i,j} \approx \Pr(Q_i = j \;|\; S = s),
\end{equation}
for $j\in\{0,1,2,3\}$. 
Following \cite{Panteleev2021degeneratequantum}, 
we define the reliability of $v_i$ as
\begin{equation}
    \gamma(v_i) = 
    \max\{p_{v_i,0},\;p_{v_i,1},\;p_{v_i,2},\;p_{v_i,3}\},
\end{equation}

Similar to its binary counterpart, Q-BP can also be improved 
by guided decimation, denoted as the quaternary belief propagation guided decimation (Q-BPGD) algorithm. 
Q-BPGD iterates between Q-BP and decimation in rounds with $r$ 
going from 1 to $n$, with early termination upon Q-BP convergence.
In each round, after running Q-BP for $T$ iterations, out of all the variable nodes not yet decimated, we pick the variable node $v_i$ with the largest reliability $\gamma(v_i)$ and decimate it by updating its channel message to
\begin{equation}
    \mu_{v_i} = \begin{cases}
        (1-\epsilon,\epsilon,\epsilon,\epsilon), & \hspace{-1mm}\text{if }
        \gamma_i^\ast = 0\\
        (\epsilon,1-\epsilon,\epsilon,\epsilon), & \hspace{-1mm}\text{if }
        \gamma_i^\ast = 1\\
        (\epsilon,\epsilon,1-\epsilon,\epsilon), & \hspace{-1mm}\text{if }
        \gamma_i^\ast = 2\\
        (\epsilon,\epsilon,\epsilon,1-\epsilon), & \hspace{-1mm}\text{if }
        \gamma_i^\ast = 3
    \end{cases}
\end{equation}
where
\begin{equation}
    \gamma_i^\ast = \underset{j\in\{0,1,2,3\}}{\arg\max}\;p_{v_i,j}.
\end{equation}

As $\epsilon$ approaches zero, this decimation effectively assigns a hard value $\widehat{q}_i$ for $Q_i$ in $\{0,1,2,3\}$ according to the bias of $v_i$.
In our software implementation for the simulation results in this section, we set $\epsilon = 1\times 10^{-10}$ for numerical stability.
The pseudo-code for the Q-BPGD algorithm over depolarizing noise is provided in Algorithm~\ref{alg:Q-BPGD}.

\begin{figure}[t]
\centering
\begin{tikzpicture}
\begin{axis}[
scale=0.95,
legend cell align={left},
legend style={
  fill opacity=0.7,
  draw opacity=1,
  text opacity=1,
  at={(1,0)},
  anchor=south east,
  draw=lightgray204,
  font=\scriptsize
},
grid=both,
label style = {font=\footnotesize},
ticklabel style = {font=\footnotesize},
grid style={darkgray176},
xmin=0.035, xmax=0.105,
xmajorgrids,
xminorgrids,
xtick={0.04,0.05,0.06,0.07,0.08,0.09,0.10},
xticklabels={0.04,0.05,0.06,0.07,0.08,0.09,0.10},
xlabel={$p$},
ymode=log,
ymin=1e-5, ymax=2e-1,
yminorgrids,
ylabel={Block Error Rate},
tick pos=left,
]

\addplot [thick, color1, mark=*]
table {%
0.10 0.144300
0.09 0.093809
0.08 0.039651
0.07 0.011972
0.06 0.003159
0.05 0.000876
0.04 0.000236
};
\addlegendentry{Q-BP}

\addplot [thick, color2, mark=*]
table {%
0.10 0.16
0.09 0.065
0.08 0.026
0.07 0.0071
0.06 0.0018
0.05 0.00028
0.04 0.000035
};
\addlegendentry{BP-OSD-10 \cite{Panteleev2021degeneratequantum}}

\addplot [thick, color4, mark=*]
table {%
0.10 0.124226
0.09 0.055219
0.08 0.021590
0.07 0.006140
0.06 0.001608
0.05 0.000295
0.04 0.000058
};
\addlegendentry{Q-BPGD, $T$ = 10}

\end{axis}
\end{tikzpicture}
\caption{Performance of various decoders on the $[[180,10,15\le d\le 18]]$ A5 code \cite{Panteleev2021degeneratequantum} over depolarizing noise. The data points of Q-BP and Q-BPGD are collected by running simulations until we observe 100 error cases. The data points of BP-OSD with order 10 are taken from \cite[Figure 6]{Panteleev2021degeneratequantum}}
\label{fig:A5_QBPGD}
\end{figure}

In Figure \ref{fig:A5_QBPGD}, we present simulation results of Q-BPGD decoder with $T=10$ on the $[[180,10,15\le d\le 18]]$ A5 code \cite{Panteleev2021degeneratequantum} over depolarizing noise. 
For comparison, Figure \ref{fig:A5_QBPGD} includes the performances of the Q-BP decoder with $T=100$ and the BP-OSD decoder with data points directly taken from \cite[Figure 6]{Panteleev2021degeneratequantum}. For the A5 code, Q-BPGD exhibits improved performance compared to Q-BP and similar performance compared to BP-OSD with order 10.

In Figure \ref{fig:B2_QBPGD}, we also present the Q-BPGD performance with $T=10$ for the $[[882, 48, 16]]$ B2 code \cite{Panteleev2021degeneratequantum}. 
Similarly, the comparison includes the Q-BP decoder with $T=100$, and the BP-OSD decoder with data points directly taken from \cite[Figure~2]{Panteleev2021degeneratequantum}. 
On the B2 code, Q-BPGD outperforms the BP-OSD decoder in the high-error regime but is surpassed by BP-OSD in the low-error-rate regime. 
We remark that besides the BP-OSD performance that we have been using as the main comparison benchmark in this paper, 
a very good decoding performance is also reported in \cite[Figure 13]{kuo2022exploiting} for the B2 code using an adaptive version of Q-BP with memory. 

\begin{figure}[t]
\centering
\begin{tikzpicture}
\begin{axis}[
scale=0.95,
legend cell align={left},
legend style={
  fill opacity=0.7,
  draw opacity=1,
  text opacity=1,
  at={(1,0)},
  anchor=south east,
  draw=lightgray204,
  font=\scriptsize
},
grid=both,
label style = {font=\footnotesize},
ticklabel style = {font=\footnotesize},
grid style={darkgray176},
xmin=0.035, xmax=0.105,
xmajorgrids,
xminorgrids,
xtick={0.04,0.05,0.06,0.07,0.08,0.09,0.10},
xticklabels={0.04,0.05,0.06,0.07,0.08,0.09,0.10},
xlabel={$p$},
ymode=log,
ymin=1e-5, ymax=2e-1,
yminorgrids,
ylabel={Block Error Rate},
tick pos=left,
]

\addplot [thick, color1, mark=*]
table {%
0.10 0.060326
0.09 0.029780
0.08 0.015077
0.07 0.006755
0.06 0.002569
0.05 0.001265
0.04 0.000406
};
\addlegendentry{Q-BP}

\addplot [thick, color2, mark=*]
table {%
0.10 0.021
0.09 0.013
0.08 0.0047
0.07 0.0017
0.06 0.00062
0.05 0.00018
0.04 0.000042
};
\addlegendentry{BP-OSD-0 \cite{Panteleev2021degeneratequantum}}

\addplot [thick, color4, mark=*,
]
table {%
0.10 0.009561
0.09 0.004651
0.08 0.001895
0.07 0.000905
0.06 0.000455
0.05 0.000210
0.04 0.000084
};
\addlegendentry{Q-BPGD, $T$ = 10}

\end{axis}
\end{tikzpicture}
\caption{Performance of various decoders on the $[[882, 48, 16]]$ B2 code \cite{Panteleev2021degeneratequantum} over depolarizing noise. The data points of Q-BP and Q-BPGD are collected by running simulations until we observe 100 error cases. The data points of BP-OSD with order 0 are taken from \cite[Figure 2]{Panteleev2021degeneratequantum}}
\label{fig:B2_QBPGD}
\end{figure}

\section{Summary}
\label{sec:summary}
In this paper, we introduce and evaluate the use of BPGD to decode QLDPC codes to encourage convergence. 
For motivation, we propose a sampling decoding approach for QLDPC codes and show how we can approximate it with BPGD.
BPGD shows strong performance compared with BP-OSD and BP-SI over bit-flip noise. 
To better understand how BPGD performs, we experiment with a randomized version of BPGD that helps illuminate the role of degeneracy in syndrome decoding.
Our experiments suggest that BPGD performed well because degeneracy allows it to achieve successful decoding along many different decimation paths.
Furthermore, we extend our guided decimation from binary BP to quaternary BP, demonstrating performance competitive compared to BP-OSD 
in the high-error regime over depolarizing noise.

\section{Funding Acknowlegement}

This material is based on work supported by the NSF under Grants 2106213 and 2120757.  Any opinions, findings, and conclusions or recommendations expressed in this material are those of the authors and do not necessarily reflect the views of the NSF.

\medskip
\bibliographystyle{quantum}
\bibliography{ref}

\appendix

\section{Proof of Theorem~\ref{thm:DQML_vs_sampling}}

The inequality $P_{\mathrm{DQML}} \le P_{\mathrm{S}}$
follows from the fact that the DQML decoder is
an optimal decoder that minimizes the error probability.

To prove $P_{\mathrm{S}} \le 2\cdot P_{\mathrm{DQML}}$, 
let's denote $E$ as a Pauli error, 
$E\mathcal{S}$ as a coset of the stabilizer group $\mathcal{S}$ in $\mathcal{P}_n\backslash \mathcal{S}$,
$S$ as a syndrome, 
and $\widehat{E\mathcal{S}}(S)$ 
as the coset with the largest posterior probability output 
by the DQML decoder given syndrome $S$.
Denote by $\openone(\cdot)$ an indicator function of an event, 
we can then write $P_{\mathrm{DQML}}$ as
\begin{equation}
\begin{aligned}
   P_{\mathrm{DQML}} 
   &= \sum_{S} \Pr(S)
   \sum_{E\mathcal{S}} 
   \Pr(E\mathcal{S}\,|\,S)
   \openone(E\mathcal{S}\neq\widehat{E\mathcal{S}}(S))\\
   &= \sum_{S} \Pr(S)\left(1 - \Pr(\widehat{E\mathcal{S}}(S)\,|\,S)\right) \\
   &= 1 - \sum_{S}\Pr(S)\cdot 
   \Pr(\widehat{E\mathcal{S}}(S)\,|\,S),
\end{aligned}
\end{equation}

Similarly, let $\widehat{E\mathcal{S}}_s(S)$ denote the 
coset picked by the sampling decoder, 
we can write $P_{\mathrm{S}}$ as
\begin{equation}
\begin{aligned}
   P_{\mathrm{S}} 
   &= \sum_{S} \Pr(S)
   \sum_{E\mathcal{S}} 
   \Pr(E\mathcal{S}\,|\,S)
   \openone(E\mathcal{S}\neq\widehat{E\mathcal{S}}_s(S))\\
   &= \sum_{S} \Pr(S)
   \sum_{E\mathcal{S}} 
   \Pr(E\mathcal{S}\,|\,S)
   (1 - \Pr(E\mathcal{S}\,|\,S))\\
   &= \sum_{S} \Pr(S)
   \left(1-\sum_{E\mathcal{S}}\Pr(E\mathcal{S}\,|\,S)^2\right)\\
   &\le \sum_{S} \Pr(S)
   \left(1-\max_{E\mathcal{S}}\Pr(E\mathcal{S}\,|\,S)^2\right)\\
   &= \sum_{S} \Pr(S)
   \left(1-\Pr(\widehat{E\mathcal{S}}\,|\,S)^2\right)\\
   &= 1 - \sum_{S}\Pr(S)\Pr(\widehat{E\mathcal{S}}\,|\,S)^2\\
   &\le 1 - \left(
   \Pr(S)\Pr(\widehat{E\mathcal{S}}\,|\,S)
   \right)^2 \\
   & = 1 - (1 - P_{\mathrm{DQML}}^2) \le 2\cdot P_{\mathrm{DQML}}.
\end{aligned}
\end{equation}
This completes the proof.

\end{document}